\documentclass[twocolumn]{jpsj2} 
\title{NMR Evidence for Antiferromagnetic Transition in the Single-component Molecular Conductor, 
[Au(tmdt)$_{2}$] at 110 K}

\author{\textsc{Yohta Hara$^{1}$, Kazuya Miyagawa$^{1,2}$, Kazushi Kanoda$^{1,2}$, 
Mina Shimamura$^{3}$, Biao Zhou$^{4}$, Akiko Kobayashi$^{3,4}$ and Hayao Kobayashi$^{2,4,5}$}} 

\inst{$^{1}$Department of Applied Physics, University of Tokyo, Bunkyo-ku, Tokyo 113-8656, Japan \\
$^{2}$ CREST-JST, Kawaguchi, Saitama 332-0012, Japan \\
$^{3}$ Research Centre for Spectrochemistry, Graduate School of 
Science, University of Tokyo, Bunkyo-ku, Tokyo 113-0033, Japan \\
$^{4}$ Department of Chemistry, College of Humanities and Sciences, Nihon 
University, Setagaya-Ku, Tokyo 156-8550, Japan \\
$^{5}$ Institute for Molecular Science, Myodaiji, Okazaki 444-8585, Japan}

\abst{We present the results of a $^{1}$H NMR study of the single-component molecular conductor, 
[Au(tmdt)$_{2}$]. 
	A steep increase in the NMR line width and a peak formation of 
the nuclear spin-lattice relaxation rate, $1/T_{1}$, were observed at around 
110 K.
	This behavior provides clear and microscopic evidences for a magnetic phase transition 
at considerably high temperature amang organic conductors.
	The observed variation in $1/T_{1}$ with respect to temperature indicates 
the highly correlated nature of the metallic phase.}

\kword{single-component molecular conductor, NMR, [Au(tmdt)$_{2}$], magnetic phase transition}

\begin{document}
\maketitle

	The majority of the molecular conductors are charge transfer salts.
	They consist of more than two types of molecules in which electrons are transferred from donor 
molecules to accepter molecules; hence, the conduction band is neither empty nor fully occupied.
	Charge transfer salts possess properties that encompass a wide variety of electronic phases 
ranging from insulator states to superconductivity \cite{Ref1}.
	It was believed that the synthesis of single-component molecular 
conductors is very difficult. 
	However, it has been recently found that the single-component molecular material 
[Ni(tmdt)$_{2}$], where tmdt denotes trimethylenetetrathiafulvalenedithiolate, 
exhibits metallic behavior for resistivity values down to a temperature limit of 0.6 K \cite{Ref2};
futhermore; \textit{de Haas van Alphen} oscillations were observed  \cite{Ref3}. 
	These observations confirm the presence of Fermi surfaces.

\begin{figure}
\begin{center}
\includegraphics[width=8.5cm]{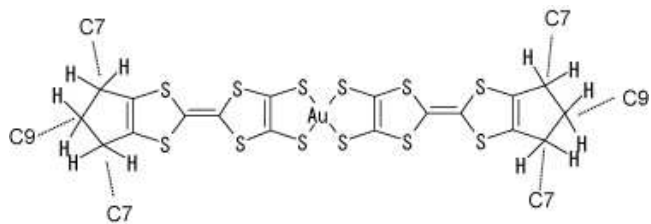}
\caption{\label{Fig1} Schematic of [Au(tmdt)$_{2}$]}
\end{center}
\end{figure}

	From the subsequent studies performed, several conductors that are analogous to single-component 
conductors were obtained\cite{Ref4}. 
	Among these conductors is the gold complex, [Au(tmdt)$_{2}$] (Figure \ref{Fig1}), 
which is isostructual with [Ni(tmdt)$_{2}$] and highly conductive 
(50 Scm$^{-1}$ at room temperature); the resistivity of the compacted pellet samples 
moderately increases with decrease in temperature, probably due to intergrain 
hopping \cite{Ref5,Ref6}.
	A very recent conductivity measurement on poly-crystals with micro-electrodes reveals their 
metallic conductivity for temperature limits down to 10 K \cite{Ref7}.
	The spin susceptibility is insensitive to temperatures above 110 K; this is 
reminiscent of the phenomenon, Pauli paramagnetism \cite{Ref5,Ref6}.
	However, below 110 K, the spin susceptibility suddenly decreases, 
and it is recovered at high magnetic fields \cite{Ref5,Ref6}.
	The ESR signal intensity suddenly decreases below 110 K \cite{Ref5}.
	Such magnetic behavior suggests an antiferromagnetic phase transition at around 110 K \cite{Ref6}.
	It is remarkable that this temperature is considerably higher than the Neel or SDW 
temperatures of charge transfer salts, for example, quasi-two-dimensional salts such as 
$\kappa$-(BEDT-TTF)$_{2}$Cu[N(CN)$_{2}$]Cl ($T_{\rm{N}}\sim $27 K) \cite{Ref8} 
and $\beta^{\prime}$-(BEDT-TTF)$_{2}$ICl$_{2}$ ($T_{\rm{N}}\sim $ 22 K) \cite{Ref9}
and quasi-one-dimensional salts such as (TMTSF)$_{2}$PF$_{6}$ ($T_{\rm{N}}\sim $ 12 K) \cite{Ref10}, 
where BEDT-TTF denotes bis(ethylenedithio)tetrathiafulvalene and 
TMTSF denotes tetramethyltetraselenafulvalene.

	The purpose of this study is to clarify the magnetic transition at the 
extraordinarily high temperature and examine the electronic state for a wide temperature range 
by means of a microscopic probe, $^{1}$H NMR.
	In this letter, we report NMR evidences obtained for the magnetic transition at 110 K.

	Since it is difficult to obtain a large single crystal of [Au(tmdt)$_{2}$], 
we used poly-crystals in this NMR study.
	The [Au(tmdt)$_{2}$] crystal is highly conductive \cite{Ref6}.
	It is well known that highly conductive materials prevent rf wave from penetrating 
into a sample (so-called skin effect).
	Hence, the samples were ground into a powder to avoid this effect. 
	The $^{1}$H NMR measurements were performed for a magnetic field of 3.66 Tesla and temperature 
ranging from room temperature down to 1.9 K.
	As shown in Fig. \ref{Fig1}, protons are located at both the edges of [Au(tmdt)$_{2}$].
	The $^{1}$H NMR spectra and the nuclear spin-lattice relaxation rate, 1/$T_{1}$, 
were measured with the so-called solid echo pulse sequence, comb$-(\pi/2)_{x}-(\pi/2)_{y}$ 
\cite{Ref11}. 
	The typical width of the $\pi$/2 pulse was 0.6 $\mu$s.
	The spectra were obtained by the Fast Fourier Transformation of the echoes.
	For low temperature ranges, where the spectra are broadened, we constructed the spectra by 
varying the resonance frequency under a constant external field in order to confirm whether 
or not the pulse width was sufficiently narrow to cover the frequency region of the entire spectra.
	We found no difference between the spectra obtained by the two methods.
	The value of 1/$T_{1}$ was measured by the standard saturation 
recovery method.
%
%
%
	The relaxation curves of $^{1}$H nuclear magnetization were well fitted to a single exponential 
function, for temperatures above 110 K.
	However, below 110 K, the relaxation curves deviated from the single exponential behavior.
	Hence, in order to characterize the curves, they were fitted to the stretched exponential 
function, exp($-(t/T_{1})^{\beta}$).

	For comparison, we also performed $^{1}$H NMR measurements on a powdered sample of 
an antiferromagnet, $\kappa$-(BEDT-TTF)$_{2}$Cu[N(CN)$_{2}$]Cl.

\begin{figure}
\begin{center}
\includegraphics[width=8.5cm]{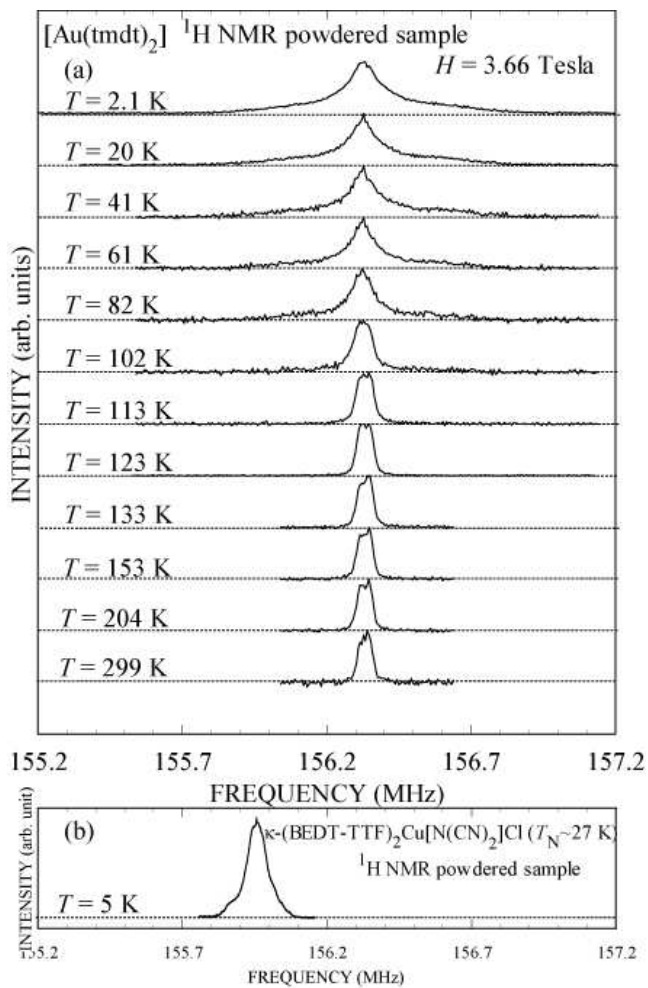}
\caption{\label{Fig2} (a) $^{1}$H NMR spectra of powdered [Au(tmdt)$_{2}$].
	(b) $^{1}$H NMR spectrum of powdered $\kappa$-(BEDT-TTF)$_{2}$Cu[N(CN)$_{2}$]Cl 
at 5 K well below $T_{\textrm{N}}$. 
The difference between the spectral centers of the two salts is due 
to the difference in their fields.}
\end{center}
\end{figure}


	The temperature-dependent $^{1}$H NMR spectra are shown in Fig. \ref{Fig2}.
	Because of the negligible contribution of the hydrogen 1$s$ orbital to the molecular orbitals 
in [Au(tmdt)$_{2}$], the Knight shift at $^{1}$H sites is too small to be resolved 
in the paramagnetic state. 
	Therefore, the line shape is governed by the temperature-independent nuclear dipole 
interactions between the $^{1}$H nuclei in the trimethylene group (Fig. \ref{Fig1}), for 
temperatures above 110 K.
	However, below 110 K, the NMR spectra are broadened.
%
%
	The line width increases with decrease in temperature;  at the lowest temperature, 1.9 K, 
the spectrum is spread over a frequency range of $\pm$ 600 kHz around the origin of shift.
	In order to characterize this line broadening, the square root of the 2nd. moment, which 
measures the spectral width, is plotted against the temperature, as shown in Fig. 3.
	The curve starts to increase at 110 K and reaches a value of 200 kHz, which is ten-times larger 
than the value at temperatures above 110 K.
	This broadening indicates the generation of local fields at $^{1}$H sites 
and provides microscopic evidence for magnetic transition at 110 K.
	As the susceptibility does not show a sudden increase below this temperature \cite{Ref5,Ref6}, 
the material cannot be ferromagnetic.
	The spectral edge of $\pm$600 kHz is considerably larger than $\pm$150 kHz for the reference material, $\kappa$-(BEDT-TTF)$_{2}$Cu[N(CN)$_{2}$]Cl, which is a commensurate antiferromagnet
 with a moment of 0.45 $\mu_{\textrm{B}}$/(BEDT-TTF)$_{2}$ \cite{Ref12}.
%
%
%
%
%

\begin{figure}
\begin{center}
\includegraphics[width=8.5cm]{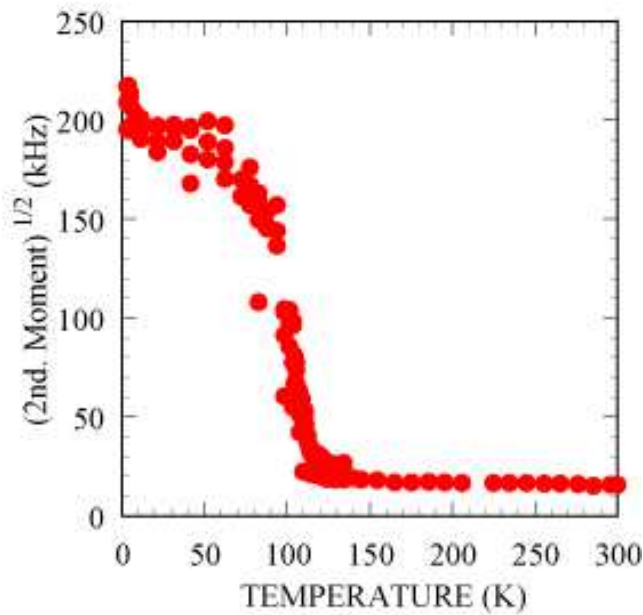}
\caption{\label{Fig3} (Color online) Temperature dependence of the square root of the 2nd moment. }
\end{center}
\end{figure}

	The NMR spectra in the ordered state appear to consist of two components.
%
%
%
%
	For the lowest temperature, fitting the spectrum to the Gaussian functions 
of its two components yields the full-width at half maximum of 120 kHz and 575 kHz. 
	The intensity ratio of the sharp and broad components is 1:2.7.
%
%
%
%
%
%
%
%
%
%
%
	The nuclear spin-spin coupling between the $^{1}$H nuclei in the trimethylene group complicates 
the echo decay curve.
	Hence, we did not perform the $T_{2}$ corrections to the intensity.

	The molecular orbital calculation predicts that the spin population on 
the $^{1}$H sites is negligible; hence, the properties of $^{1}$H NMR spectra 
below $T_{\rm{N}}$ are governed by the dipole fields from electron spins on other atomic sites.
	Moreover, considering that the external field of 3.66 Tesla in the present experiment 
is larger than the spin-flop field estimated from the field dependence of the spin susceptibility, 
\cite{Ref6} the on-site $^{1}$H hyperfine field, if any, is not expected to affect 
the NMR spectra because the $^{1}$H NMR spectra reflect the parallel component of the local field to 
the external field.

\begin{figure}
\begin{center}
\includegraphics[width=5cm]{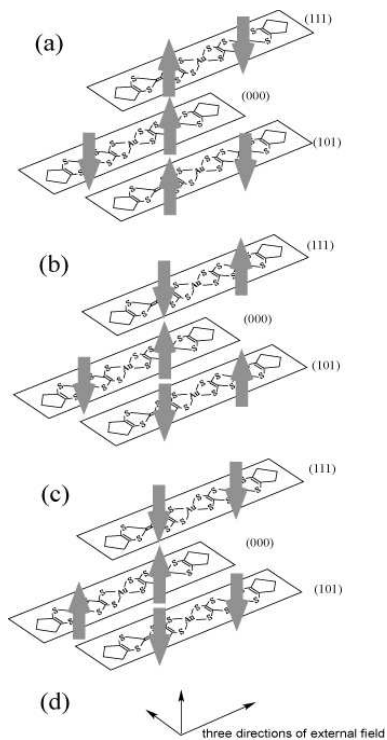}
\caption{\label{Fig4} Spin structures assumed in the calculations. 
Pattern (a) is suggested by the first principle 
band calculation.$^{13}$
(d) shows the directions of the external field assumed in the calculations}
\end{center}
\end{figure}

%
	As the spin structure and easy axis are not determined experimentally, 
we assume three possible antiferromagnetic spin arrangements (shown in Fig. \ref{Fig4}) 
for the calculations of the local field.
	It must be noted that pattern (a) is suggested by the first-principle calculation \cite{Ref13} 
and the Hubbard model calculation \cite{Ref13b}. 
	The intramolecular antiferromagnetic spin configuration ((a) and (b)) is 
among the candidates because the two tmdt molecular orbitals in [Au(tmdt)$_{2}$] interact weakly. 
%
%
%
%
%
	The interplane spacing between the [Au(tmdt)$_{2}$] molecules located on the 
(000), (111) and (101) positions is  approximately 3.5 \AA \cite{Ref5}. 
%
%
%
%
	Because this distance is shorter than the longest length of tmdt, the local fields from two neighboring molecules are considered along with that from the on-site molecule. 
%
%
%
	We ignore the local field from Au, if any, which is farther from the methylene group 
as compared to sulfurs and carbons in tmdt.
%
%
	Based on these assumptions, we calculated the local fields at two $^{1}$H sites.
	One site (we call H1 proton) connects to C7 carbon and the other site (H2 proton) connects 
to C9 carbon.

%
\begin{figure}
\begin{center}
\includegraphics[width=8.5cm]{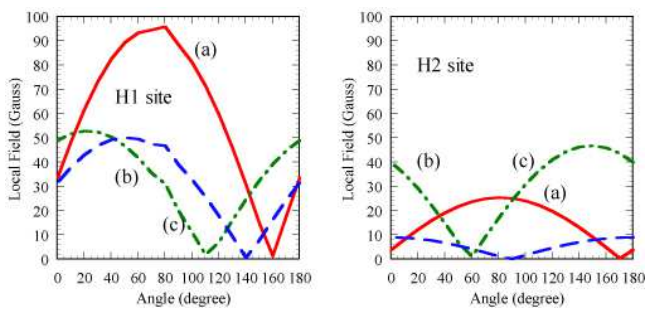}
\caption{\label{Fig5} (Color online) 
Parallel component of the local field. (a), (b), and (c) correspond 
to the spin structures shown in Fig.4. 
The external field is parallel to the longest direction of the molecule. 
The moment is rotated within the plane perpendicular to the external 
field.}
\end{center}
\end{figure}

	Figure \ref{Fig5} shows the profile of the parallel component of the local fields, 
for the rotation of a moment of 0.5  $\mu_{\textrm{B}}$ on tmdt within a plane perpendicular 
to the external field, which is assumed parallel to the long axis of the molecule.
	The solid, dash-doted, and  dashed lines represent the local fields in patterns 
(a), (b) and (c), respectively.
%
%
%
%
	Because this experiment is performed for a powdered sample in which the external field is 
directed arbitrarily with respect to the constituent microcrystals, we also calculate the local 
fields along two other directions of the external field  (see Fig. \ref{Fig4}).
	In every situation, the local field at the H1 site is 1 $\sim$ 5 times larger than that at 
the H2 site.
%
%
%
	This may explain the double structure of the observed spectra.
	In every case, the maximum value at C7 site is 100 Gauss in pattern (a), 
60 Gauss in pattern (b), 60 Gauss in pattern (c), which correspond to 425 kHz, 250 kHz, 
and 250 kHz in the $^{1}$H NMR spectral edge at the H1 site, respectively.
	By comparing these values with the observed spectral edge of 600 kHz, 
we estimate the local moment at 0.7$\sim$ 1.2 $\mu_{\textrm B}$/tmdt. 
	The [Au(tmdt)$_{2}$] contains an odd number of electrons. 
	This indicates that a $S$=1/2 spin generally remains in a molecule, unless an extraordinary situation exists, for example, situations in which three or more orbitals in [Au(tmdt)$_{2}$] are 
degenerate.
	Therefore, in the likely case of symmetrical spin distribution in [Au(tmdt)$_{2}$], the maximum 
value of the moment on tmdt is 0.5 $\mu_{\rm{B}}$.
	The evaluated moment value exceeds this upper limit.
%
%
	The calculations show that the local fields at the $^{1}$H site are dominated by the on-site 
molecular spin, and they are sensitive to the spin density of the C7 and C9 sites. 
	Although a very small spin population on both the carbon sites (C7 and C9 sites) is predicted 
by molecular orbital calculations, the local field at the H1 site increases by 10 $\sim$ 15 gauss 
if a spin population of 0.6 \% exists on the C7 site. 
	The local fields from the neighboring molecules are smaller than that from the on-site molecule; 
however, they are not negligible.
	Moreover, the absolute value and sign of the local fields from the neighboring molecules are 
sensitive to the position of the proton.
	This can result in the overestimation of the local moment.

	The estimated value should not be considered as it is; however, 
the value suggests that the moment of the present system is larger than 
0.08 $\mu_{\textrm{B}}$ of the spin-density-wave (SDW) salt, 
(TMTSF)$_{2}$PF$_{6}$ \cite{Ref10} and 0.45 $\mu_{\textrm{B}}$/(BEDT-TTF)$_{2}$ 
(0.23 $\mu_{\textrm{B}}$ per BEDT-TTF) of 
the Mott-insulating salt $\kappa$-(BEDT-TTF)$_{2}$Cu[N(CN)$_{2}$]Cl \cite{Ref12}.

\begin{figure}
\begin{center}
\includegraphics[width=8.5cm]{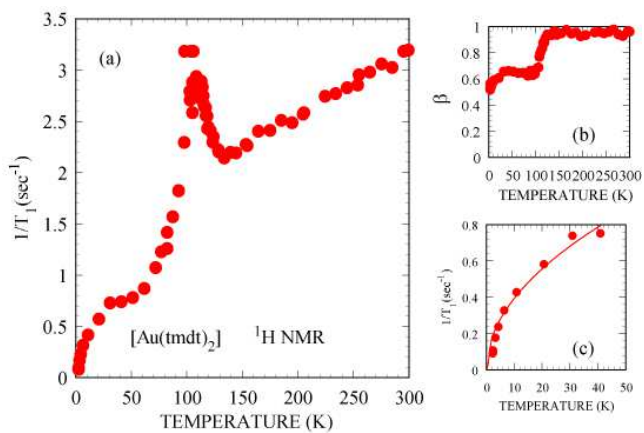}
\caption{\label{Fig6} (Color online) (a)Temperature dependence of $1/T_{1}$.
	Above 110 K, $1/T_{1}$ is unambiguously determined from a simple exponential recovery curve of 
nuclear magnetization.
	Below 110 K, $1/T_{1}$ is determined by a stretched exponential function, 
exp[-(t/$T_{1}$)$^{\beta}$], fitting the non-single exponential recovery. 
	(b) Temperature dependence of fitting parameter $\beta$. 
	(c) Temperature dependence of $1/T_{1}$ at low temperatures. 
The solid line expresses $T^{0.5}$.}
\end{center}
\end{figure}

	The temperature dependence of $1/T_{1}$, is shown in Fig. 
\ref{Fig6}.
	The absolute value of $1/T_{1}$ at room temperature, 3 s$^{-1}$, is one order of magnitude 
larger than the values of the organic superconductors, $\kappa$-(BEDT-TTF)$_{2}$X [X=Cu(NCS)$_{2}$ 
\cite{Ref14} and Cu[N(CN)$_{2}$]Br \cite{Ref15}]\cite{Ref16}, 
which are located near the Mott transition \cite{Ref17}.
	The $1/T_{1}$ values decrease with temperature from 300 K; and they exhibit 
gradual temperature dependence as compared to the Korringa relation, 
$1/T_{1} \propto T$, which should be followed by uncorrelated metallic electrons 
with temperature-insensitive susceptibility. 
%
%
%
%
%
	The $1/T_{1}$ values increase below 130 K and shows a peak at 110 K, where the line 
broadening is observed.
	The peak formation is the manifestation of the magnetic critical slowing down, and 
it provides further evidence for the magnetic phase transition at 110 K.
%
%
%
%
%
	Below 110 K, $1/T_{1}$ decreases rapidly.
	Around 30 $\sim$ 50 K, $1/T_{1}$ shows a broad shoulder, followed by a rapid decrease again 
below 30 K.
	The nuclear relaxation curve becomes inhomogeneous below the magnetic transition, 
as shown in Fig. \ref{Fig6}(b), which shows the temperature dependence of the exponent, 
$\beta$, in the stretched exponential function, 
exp(-(t/$T_{1}$)$^{\beta}$), fitting the relaxation curve.
	The fitting parameter, $\beta$, is related to the inhomogeneity of $1/T_{1}$ 
at the $^{1}$H sites.
	A clear deviation in the $\beta$ values from 1.0, below 110 K, originates from 
the emergence of the inhomogeneous local field associated with the antiferromagnetic ordering.
	Below 30 K, $\beta$ shows a moderate decrease toward 0.5; this likely indicates 
a further increase in the antiferromagnetic moment.
	These observations consistent with the further line broadening 
below 30 K (Fig \ref{Fig3}). 
	Each proton may have different hyperfine tensors; hence, the relaxation curve is expected 
to be non-single exponential, even above $T_{\textrm {N}}$. 
	One possible explanation for the single exponential relaxation curve 
above $T_{\textrm{N}}$ can be provided in terms of the $T_{2}$ process, which averages the values of 
1/$T_{1}$ at different $^{1}$H sites, above $T_{\textrm{N}}$, where the nuclear-dipolar fields 
from the adjacent $^{1}$H nuclei in trimetylene groups are considerably larger than 
the distribution of hyperfine fields.
%
%
%
%
%

%
	In the general case of antiferromagnetic transition, 
$1/T_{1}$ shows a power law or gapped behavior below the transition temperature, 
reflecting spin-wave excitations.
	However, in this case, $1/T_{1}$ levels off at around 50 K and 
shows an anomalous decrease that is roughly characterized by $1/T_{1} \propto T^{0.5}$ 
for temperatures between 3 and 40 K (Fig. \ref{Fig6}(c)), thereby suggesting the additional 
relaxation contribution besides the spin wave below 110 K.
	The present system is metallic in the entire temperature range studied \cite{Ref7}; 
however, it should be regarded as a highly correlated metal with magnetic ordering and 
independent of the Korringa law in $1/T_{1}$ for any temperature range. 
	It must be noted that the $1/T_{1}$ behavior 
is not described by the SCR theory of the weak itinerant antiferromagnetic metal \cite{Ref17b}, 
which predicts that $1/T_{1} \propto T/\sqrt{T-T_{\rm N}}$ for $T \geq T_{\rm N}$ and 
$1/T_{1}\propto T$ for $T \ll T_{\rm N}$; the expression $T/\sqrt{T-T_{\rm N}}$ fails to fit 
the $1/T_{1}$ data above $T_{\rm N}$, and the low-temperature behavior of $1/T_{1} 
\sim T^{0.5}$ contradicts the SCR expectation of $1/T_{1} \sim T$.
%

	The first principles band-structure calculations predict a part of the Fermi surfaces with 
a good nesting vector along $a^{\ast}/2$ \cite{Ref13}, which favors the imperfect nesting. 
	This explains the survival of the metallic state after antiferromagnetic transition, as observed 
in the Cr system \cite{Ref18}.
%
%
%
%
	Further study will be necessary to discuss the similarities and differences between 
[Au(tmdt)$_{2}$] and Cr systems.
%
%
%
%
%
%
%
%
%
%
	It is well known that low-dimensionality depresses a phase transition and causes spin 
contraction due to quantum fluctuations.
	Unlike the majority of charge transfer salts, the [Au(tmdt)$_{2}$] compound is 
a three-dimensional system, as shown by the Fermi surface topology \cite{Ref13}.
	The temperature region where the critical slowing down is observed is 
narrower than that of the quasi-two dimensional system 
$\kappa$-(BEDT-TTF)$_{2}$Cu[N(CN)$_{2}$]Cl \cite{Ref8}.
	It is probable that three dimensionality causes the high transition temperature and 
the large ordered moment.

	In summary, the magnetism of the single-component molecular conductor, [Au(tmdt)$_{2}$], 
has been investigated by means of $^{1}$H NMR.
	The spectral broadening and relaxation-rate enhancement at 110 K provide microscopic 
evidence to prove that [Au(tmdt)$_{2}$] undergoes a magnetic phase transition at extraordinarily 
high temperature among organic conductors.
	The magnitude and temperature dependence of relaxation rate indicates that [Au(tmdt)$_{2}$] is 
an unconventional metal with an antiferromagnetic order and anomalous $1/T_{1}$ behavior at 
low temperatures.
%

	The authors thank S. Ishibashi, H. Seo, H. Fukuyama, Y. Okano for their helpful discussions.
	This study is supported by MEXT KAKENHI in Priority Areas of 
``Molecular Conductors'' (No. 15073204) and ``Physics of New Quantum Phases in 
Superclean Materials'' (No.17071003), 
and JSPS KAKENHI (No. 15104006). One of the authors (K.M.) is indebted to the Asahi 
Glass Foundation for financial supports.


\begin{thebibliography}{999}
\bibitem{Ref16} In these compounds, the $^{1}$H-NMR $1/T_{1}$ is dominated by the thermal 
vibration of ethylene groups in the molecule around room temperature.
With analysis to subtract this contribution to $1/T_{1}$, the electron contribution to $1/T_{1}$ is 
estimated at $\sim$0.3 s$^{-1}$ at room temperature.
\end{thebibliography}
\end{document}